\documentstyle{mn}
\input psfig.tex

\newif\ifAMStwofonts
\newcommand{\ltsima} {$\; \buildrel < \over \sim \;$}
\newcommand{\gtsima} {$\; \buildrel > \over \sim \;$}
\newcommand{\lta} {\lower.5ex\hbox{\ltsima}}
\newcommand{\gta} {\lower.5ex\hbox{\gtsima}}

\title{The BL Lac heart of Centaurus A}
\author[M. Chiaberge et al.]
       {M. Chiaberge$^1$, A. Capetti$^2$, A. Celotti$^1$\\
        $^1$SISSA/ISAS, Via Beirut 2-4, I-34014 Trieste, Italy. chiab@sissa.it, celotti@sissa.it\\
	$^2$Osservatorio Astronomico di Torino, Strada Osservatorio 20, I-10025
	Pino Torinese, Italy. capetti@to.astro.it}
\date{Accepted ...
      Received ...;
      in original form ...}

\pagerange{\pageref{firstpage}--\pageref{lastpage}}
\pubyear{2000}

\begin{document}

\maketitle

\label{firstpage}

\begin{abstract}

Emission from the nucleus of the closest radio galaxy, Centaurus A, is
observed from the radio to the gamma ray band. We build, for the first
time, its overall Spectral Energy Distribution (SED) that appears to
be intriguingly similar to those of blazars, showing two broad peaks
located in the far-infrared band and at $\sim 0.1$ MeV respectively.
The whole nuclear emission of Centaurus A is successfully reproduced
with a synchrotron self-Compton model.  The estimated physical
parameters of the emitting source are similar to those of BL Lacs,
except for a much smaller beaming factor, as qualitatively expected
when a relativistic jet is orientated at a large angle to the line of
sight.
These results represent strong evidence that Centaurus A is indeed a
misoriented BL Lac and provide strong support in favour of the
unification scheme for low luminosity radio-loud AGNs.  Modeling of
the SED of Centaurus A also provides further and independent
indications of the presence of velocity structures in sub-pc scale
jets.

\end{abstract}


\section{Introduction}
\label{intro}

Unification models for radio-loud active galactic nuclei (AGN) are
usually tested by comparing the extended properties (e.g. radio and optical
emission, environment) of the beamed sources and their putative parent
population.  Clearly, more compelling results could be obtained
by directly considering the nuclear AGN emission.  

The optical nuclear emission has been recently 
identified in FR~I radio-galaxies through HST observations of a complete
subsample of 3CR sources (Chiaberge, Capetti \& Celotti 1999). 
The presence of a strong correlation between radio and optical
emission provides a straightforward indication of a direct association 
between radio cores and the faint central compact cores observed by HST,
arguing for a common non-thermal synchrotron origin of
both components.  
This discovery, by itself, provides qualitative support for
the low-luminosity radio loud AGN unification models 
(Barthel 1989;
Urry \& Padovani 1995 for a review). 
In fact, in the frame
of the unification schemes, the non-thermal beamed emission which
dominates in blazars, should also be present in radio galaxies,
although strongly de-amplified.

Nonetheless, a more quantitative
analysis showed that cores of radio galaxies
are largely overluminous with respect to what is expected from
mis-oriented BL Lac jets, for typical values of the bulk Lorentz
factor (Chiaberge et al. 2000).  
In order to reconcile this result with the unification
scheme, velocity structures in the jet have been suggested, where a
fast spine is surrounded by a slower (but still relativistic) layer.

Modeling of SED could provide an even more stringent test for the
unified scheme, as it represents a fundamental tool for obtaining
information on the physical conditions of the emitting region, as it
is usually done for blazars (e.g. Ghisellini et al. 1998).
Unfortunately, the nuclear SED of radio-galaxies are in general not
sufficiently well sampled to follow directly this approach.  More
information on their nuclear emission, in addition to the radio and
optical data, comes from X-ray
observations, but the available data are still unsuited to deriving
strong constraints (see e.g. Capetti et al. 2000).

However, there is one radio-galaxy for which a sufficiently large set
of photometric data is available to build a well sampled SED: the
nearest radio galaxy Centaurus A. The presence of an AGN in Centaurus
A is evident from the radio through the $\gamma$--ray band, as
detailed in the following section. In this Letter we take advantage of
this broad band energy coverage to gain a better understanding of the
nature of the nuclear emission of Centaurus A.  The paper is
structured as follows: in Sect. 2 we build the nuclear SED, which is
modeled in the frame of a pure synchrotron self-Compton scenario in
Sect. 3; finally, in Sect. 4 we examine and discuss our findings in
the light of the unification schemes.

We adopt a distance to Centaurus A of 3.5 Mpc (Hui et al. 1993).

\section{The nuclear Spectral Energy Distribution}
\label{dati}

In order to build the overall nuclear SED of Centaurus A, we collected
from the literature the nuclear fluxes spanning the widest range of
frequencies, from the radio to the gamma-ray band.

Starting from the highest energies, Centaurus A is at present the only
radio galaxy detected in the $\gamma$--ray band by CGRO. 
When it was observed for two weeks in 1991, during an intermediate
state of activity, the spectra obtained by OSSE and COMPTEL
(and the EGRET flux) appear to be smoothly connected, forming a well
defined emission peak at $\sim 0.1$ MeV, in a $\log(\nu F_{\nu})$ vs
$\log\nu$ representation (Steinle et al. 1998). These are the only
published simultaneous data in such bands. 
In the X--ray band an absorbed ($N_{\rm H} \sim
10^{23}$ cm$^{-2}$) power--law component with energy index $\alpha
\sim 0.7 - 0.9$ ($F_\nu \propto \nu^{-\alpha}$) 
has been detected by both RXTE and $Beppo$SAX
(MECS+PDS data are reported here) (Rothschild et al. 1999, Grandi et
al. 2000). The X-ray flux is also variable (more than a factor of $4$)
on timescales of a few hours.
A significant improvement in our knowledge of the low energy part of
the spectrum (mm-optical) has been made thanks to HST, ISOCAM and SCUBA
observations. In particular, the
near-IR nuclear component is clearly seen in the HST/NICMOS images, allowing 
Marconi et al. (2000)
to accurately determine the position of the near IR nucleus and
thus the identification of a faint optical counterpart in the WFPC2
I and V bands images.
The ISO satellite has provided
a spectrum of the nuclear component in the range 5 - 18 $\mu$m 
(Mirabel et al. 1999) whose continuum we fitted with a power law. 
From the same work and from Hawarden et al. (1993)
we derived the sub-mm SCUBA measurements.   
The radio band is covered by VLA (Burns et al. 1983) and VLBI
(Tingay et al. 1998) observations.  

\begin{figure}
\psfig{file=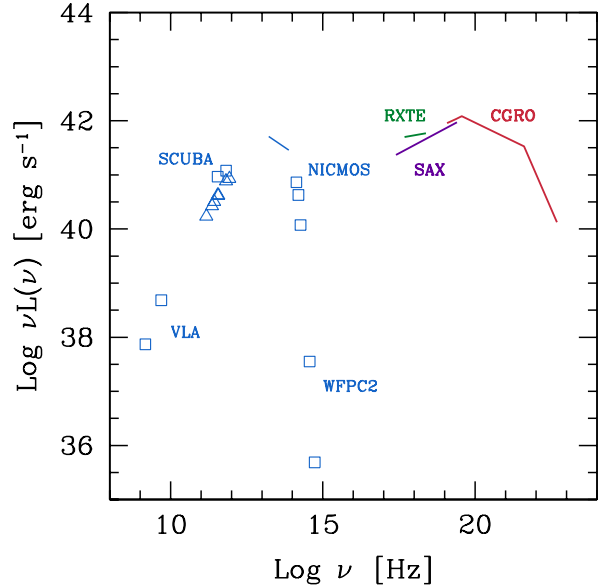,width=8.5truecm,height=8.5truecm}
\caption{The observed spectral energy distribution of the nucleus of
Centaurus A (see text).}
\label{sed_senza}
\end{figure}

These data produce the observed SED shown in Fig. \ref{sed_senza} in a
$\log(\nu L(\nu)) - \log \nu$ representation.
The SED appears to be composed by two broad peaks: the low energy one
reaching its maximum in the far-IR, between $10^{12}$ and $10^{13}$ Hz
and the high energy one with a well defined maximum at about 0.1
MeV. As both maxima occur in spectral bands essentially unaffected by
dust/gas absorption, the double-peaked shape is well established
regardless of extinction.  The SED of Centaurus A thus intriguingly
resembles that of blazars and indeed (qualitatively) what expected in
a misaligned BL Lac object.  In fact, according to the unified
scenarios for jetted AGN, the counterpart of Centaurus A when seen at
small angle with the jet axis should appear as a BL Lac.
This represents a strong clue that Centaurus A nuclear emission might be
interpreted by analogy with blazars, whose low energy peak is
attributed to non-thermal synchrotron emission and the high energy one
to inverse Compton scattering of softer photons. 

We will explore this possibility in the next section, modeling the
observed SED in order to obtain information on the physical parameters
of the source.

\section{The synchrotron self-Compton scenario}

The SED of blazars is interpreted as due to synchrotron
and inverse Compton emission. In this frame, there are two main
possibilities for the nature of the soft photons which are scattered
to high energies: this radiation field can be either internal to the
source, i.e. the synchrotron emission itself (synchrotron self-Compton,
SSC, Maraschi, Ghisellini \& Celotti 1992, Bloom \& Marscher 1993) or
external, e.g. photons emitted by either the accretion disc or the Broad Line
Region (Dermer \& Schlickeiser 1993, Sikora,
Begelman, Rees 1994, Ghisellini \& Madau 1996).
For the lowest luminosity blazars (i.e. BL Lacs) the contribution of
the external photon field is probably small (Ghisellini et
al. 1998). In fact the lack of strong broad emission lines (Marconi et
al. 2001 for Centaurus A), disc and dust emission in their spectra
are typical of these sources and argues against substantial photons
fields external to the jet.  This suggests the interpretation of the observed
nuclear emission as pure SSC radiation from the {\it BL Lac heart} of
Centaurus A.

\subsection{The model}
\label{themodel}

The simplest representation of the source is a spherical
homogeneous region, embedded in a tangled magnetic field.
Relativistic electrons are continuously injected at a rate $Q(\gamma)$
[cm$^{-3}$ s$^{-1}$] $\propto \gamma^{-p}$ between $\gamma_{\rm min}$
and $\gamma_{\rm max}$ ($\gamma$ being the Lorentz factor), and they
loose their energy radiatively.
In this scenario, 
the free parameters of the model are: the
size of the source $R$, the magnetic field $B$, the injected
luminosity $L_{\rm inj}$, the relativistic beaming factor
$\delta=[\Gamma(1-\beta \cos\theta)]^{-1}$ (where $\theta$ is the
angle between the jet axis and the line of sight), $\gamma_{\rm min}$,
$\gamma_{\rm max}$ and the slope $p$. The resulting electron
distribution at equilibrium is a broken power-law. 
For details, see Ghisellini et al. (1998) and Chiaberge \&
Ghisellini (1999).

When the Klein-Nishina decline of the cross section for Compton
scattering and pair production effects can be neglected, the physical
parameters of the source are well constrained once the frequency and
the intensity of the peaks of the SED are known and an estimate of the
variability time scale is available. As a first step, let us
analytically evaluate the physical parameters.  We can evaluate the
magnetic field $B$, the beaming factor $\delta$, and $\gamma_{\rm b}$,
the energy of the electrons which emit radiation at the peak
frequencies\footnote{We remind that $\gamma_{\rm b}$ corresponds to
$\gamma_{\rm min}$ if the slope of the injected electron distribution
is $p>2$; otherwise it lies between $\gamma_{\rm min}$ and
$\gamma_{\rm max}$ (Ghisellini et al. 1998).}, using simple
expressions relating them to the observed quantities (e.g.
Ghisellini, Maraschi \& Dondi 1996).
In the case of Centaurus A, the position of the high energy peak is
set at $\nu_{\rm c} \sim 3 \times 10^{19}$ Hz, while the low energy
one, less constrained by the observations as it lies in region
without photometric measurements, is around $\nu_{\rm s} \sim
10^{13}$ Hz.  Adopting such values, we can derive 
$\gamma_{\rm b}$ and the product $B \times \delta$:
\begin{displaymath}
\gamma_{\rm b} = \left(\frac{3}{4} 
\frac{\nu_{\rm c}} {\nu_{\rm s}}\right)^{1/2} = 1.5 \times 10^3
\end{displaymath}
\begin{displaymath}
B \delta = \frac{\nu_{\rm s}}{3.7 \times 10^{6} \nu_{\rm c}} = 0.9\, {\rm G} .
\end{displaymath}
The values of $B$ and $\delta$ can be disentangled by using
the observed peak luminosities (in units of $10^{45}$ erg s$^{-1}$)
and the variability timescale\footnote{The minimum variability time
scale of $\sim 1$ day is determined from the OSSE observations (Kinzer
et al. 1995), although shorter time scales variations have been
observed.}  $t_{\rm d} \sim 1$ (in days) as additional constraints, i.e.:
\begin{displaymath}
\delta = 1.67 \times 10^{4} \left( \frac{\nu_{\rm c}}{\nu_{\rm s}^2 t_{\rm d}} \right)^{1/2}  
            \left(  \frac{L^2_{\rm C,45}}{L_{\rm s,45}} \right)^{1/4} \sim 1.6 
\end{displaymath}
where $t_{\rm d}=1$, $L_{\rm s,45}\sim L_{\rm C,45}=10^{-3}$ are used, and
\begin{displaymath}
B = 2.14 \times 10^{-11} \frac{\nu^3_{\rm s} t_{\rm d}^{1/2}}{\nu_{\rm c}^{3/2}} \left( 
              \frac{L_{\rm C,45}}{L_{\rm s,45}^2} \right)^{1/4} \sim 0.7\, {\rm G}
\end{displaymath}

A more accurate approach is to obtain the equilibrium solution of the
continuity equation which governs the temporal evolution of the
emitting electron distribution $N(\gamma,t)$, also taking into account
the effects of the Klein-Nishina decline and the possible escape of
particles from the source on a timescale $t_{\rm esc}$:
\begin{displaymath}
\frac{\partial N(\gamma,t)}{\partial t} = \frac{\partial}{\partial\gamma} 
\left[ \dot\gamma(\gamma,t) N(\gamma,t)\right] + Q(\gamma,t) - 
\frac{N(\gamma,t)}{t_{esc}}=0
\end{displaymath}
where $\dot\gamma= \dot\gamma_{\rm s} + \dot\gamma_{\rm C}$ is the
total (synchrotron + self--Compton) cooling rate.  We solve the
equation numerically using the code extensively described by Chiaberge
\& Ghisellini (1999), which also calculates the spectrum emitted by
the resulting electron distribution.  A significant advantage of this
method is that it produces a continuous representation of the SED
which can be directly compared with the observations.

\begin{figure}
\psfig{file=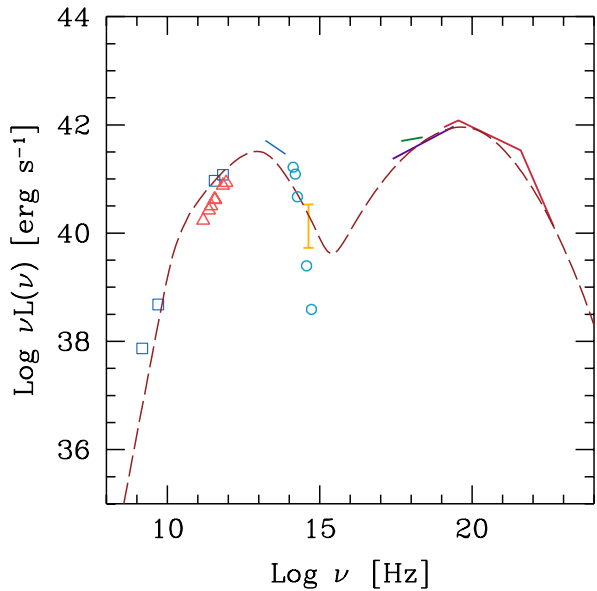,width=8.5truecm,height=8.5truecm}
\caption{The nuclear SED of Centaurus A.  The empty circles represent
the optical and near-IR luminosities de-reddened with $A_V=8$. The
dashed line is the prediction of the SSC model described in the
text. The large error bar represents the
optical luminosity predicted using the radio-optical core flux
correlation for FR~I  found by Chiaberge et al. (1999).}
\label{fig2}
\end{figure}

Before proceeding, we must consider the effects of
the prominent kpc--scale dust lane of Centaurus A that causes
significant extinction, particularly in the optical and near-IR bands.
Marconi et al. (2000) found color excesses as high as E(B-V)
$\sim$ 2.5 in the circumnuclear regions which can be considered as a
lower limit to the nuclear absorption. We thus de-reddened the
observed data points using $A_V \sim 8$ and adopting the reddening law
of Cardelli et al. (1989).  The main effect of this
procedure is to substantially reduce the depth of the SED minimum at
$\nu \sim 10^{15}$ Hz, as can be seen in the corresponding corrected
SED presented in Fig. \ref{fig2}.

The overall agreement between model and data is quite remarkable. The
``fit'' to the SED of Centaurus A obtained with the SSC model is
represented as the dashed line in the same Fig. \ref{fig2}, and the
corresponding model parameters are listed in Table \ref{param}.  We
point out that the optical core luminosity predicted by the
radio-optical correlation valid for FR~I radio galaxies (Chiaberge et
al. 1999) is consistent with $A_V = 10-13$, slightly higher than the
adopted value.  Using this $A_V$, the model would reproduce the
observed SED even more closely.

\section{Discussion}

All of the physical parameters inferred from the model (see Table
\ref{param}) are in the range spanned by blazars (e.g. Ghisellini et
al. 1998), except for the beaming factor, which is significantly
lower.  A smaller amount of beaming is indeed qualitatively expected
in the frame of the unification schemes for AGN, as radio galaxies are
believed to be observed at angles with the jet axis larger than blazars.  
In the case of Centaurus A, this is directly
confirmed by VLBI observations which, in order to account for the
observed jet-counterjet ratio, imply an
angle between $50^{\circ}$ and $80^{\circ}$ (Tingay et al. 1998).  
The fact that nuclear
emission of Centaurus A can be successfully modeled as SSC radiation
with physical parameters consistent with those used for BL Lacs and a
lower value of $\delta$ provides evidence that the source is
indeed a misoriented BL Lac and thus supports the unification scheme
for low luminosity radio-loud AGNs.

\begin{table}
\caption{Model parameters for the SED of Centaurus A}
\label{param}
%\begin{tabular}{@{}ll}
\begin{tabular}{|llll|}

$R$             &   $1.2\times 10^{16}$ cm           & $p$             &     3.0            \\   
$B$             &      0.5\, G                       & $\gamma_{min}$  &   $2\times 10^3$   \\              
$\delta$        &      1.2                           & $\gamma_{max}$  &   $1\times 10^4$   \\                  
$L_{inj}$       &   $2.7\times 10^{42}$ erg s$^{-1}$ & $t_{esc}$       &   $10 \times R/c$  \\

\end{tabular}

\medskip 
$L_{\rm inj}$ is
the injected power (in particles of energy equal to $\gamma m_{\rm e} c^2$).

\end{table}

However the bulk Lorentz factors inferred for BL Lacs are typically in
the range $\Gamma \sim 15-20$. For such a $\Gamma$, at the inferred
angles of sight, an even lower value of $\delta$ would be
expected. More quantitatively, for the above range of viewing angles,
such velocities imply a $\delta$ factor between 0.18 and 0.06, which
are incompatible with the values found from the (analytic and model
fitting) estimates.  On the other hand the SSC model strongly
constrains the value of $\delta$, as it depends only on parameters
which are observationally well determined: i) the variability
timescale (a value lower than the assumed one would result in a higher
$\delta$); ii) the position and (weakly) the luminosity of the
emission peaks, which are well defined unless (unobserved) dramatic
variability has affected the measurements (variations by a factor of
$\sim 2-4$ of do not significantly change $\delta$).  From the
analytic relations reported in Sect. \ref{themodel} it turns out that
it cannot be significatively smaller than $\sim 1$. Furthermore, based
purely on geometrical considerations, at large viewing angles the
beaming factor cannot be higher than $\delta \sim 1.3$ (even for
$\theta=50^{\circ}$ and independently on the value of $\Gamma$). The
highest value of $\delta$ is reached for $\Gamma\sim \delta$ while for
higher Lorentz factors $\delta$ rapidly decreases.  Therefore, our
results constrain the Doppler factor of the observed region to be
$\delta \sim 1$ and this in turn sets an upper bound to the Lorentz
factor, $\Gamma \lta 3-5$.  These values are consistent with the
mildly relativistic proper motions observed on sub pc-scales (Tingay
et al. 1998).

As already discussed in the Introduction, Chiaberge et al. (2000)
found a similar indication that the emission in FR~I cores is not
quantitatively compatible with being mis-oriented high Lorentz factor
($\Gamma \sim 15-20$) BL Lac jets. In fact, cores of radio galaxies
are overluminous (both in the optical and radio band) by factors
$10-10^5$ with respect to the predictions of a ``one velocity beaming
model''
for BL Lacs with similar extended radio power.  Within the unification
scenario the simplest and rather plausible hypothesis to account for
such discrepancy is to assume a structure in the jet velocity field in
which a fast spine is surrounded by a slow layer. The slower component
has still to be relativistic and we found that for
$\Gamma_{\rm layer}\sim 2$ we can account for the unification of FR~I
cores and BL Lacs.  The observed emission in aligned objects would
then originate mostly from the jet spine, while in the misaligned
ones would be dominated by emission from the slower layer.  In this
hypothesis, the radiation observed from the nucleus of Centaurus A can
be identified with emission from the jet layer.  Modeling the SED
of Centaurus A thus provides an independent indication of
the presence of velocity distribution in sub-pc scales jets.

\subsection{Centaurus A and the blazar sequence}

The population of blazars shows a link between the overall spectral
properties and the nuclear radio luminosity (Fossati et al. 1998,
Ghisellini et al. 1998): an increase in the nuclear radio luminosity
(at 5 GHz) corresponds to an increase in the bolometric luminosity,
the ratio between inverse Compton and synchrotron luminosities and a
decrease in the peak frequencies. This reflects the progressive trend
from FSRQ, to High and Low Energy Peaked BL Lacs (HBL and LBL,
respectively; Padovani \& Giommi 1995).  Furthermore, the extended
luminosity of HBL and LBL differs, on average, by a factor $\sim$ 100,
probably due to the (although weak) correlation between the nuclear
and extended radio emission (e.g. Giovannini et al. 1988).

In order to determine if/how Centaurus A fits in this blazar sequence
we ``beamed'' its SED with a typical BL Lac Doppler factor,
$\delta_{\rm BL} \sim 15$, starting from our fiducial value $\delta
\sim 1$. Clearly the effect of this procedure is to shift the observed
SED by a factor of $\delta_{\rm BL}$ in frequency and $\delta_{\rm
BL}^{a}$ in luminosity ($a=3$ and $a=4$ in the case of emission from a
continuous jet or an emitting sphere, respectively).
The resulting peak
frequencies are $\log \nu_{\rm s} \sim 14$ and $\log \nu_{\rm c} \sim
20.5$ and their luminosities are in the range $\log \nu L_\nu = 45-46$
and $\log \nu L_\nu = 45.5 - 46.5$, for the synchrotron and inverse
Compton components respectively.  Although this procedure has large
uncertainties, the inferred values are typical of LBL.  This also
appears to be consistent with the fact that the extended radio power
of Centaurus A ($ L_{\rm 408 MHz} \sim 10^{31.5}$ erg s$^{-1}$) is also in
the range spanned by LBL, although close to its lower end.
Therefore, although velocity structures appear to be
present in the jet, 
it is plausible that the physical conditions in the external
layer might be similar to those of the spine. Clearly, it is very
important to further investigate this issue by considering a large
sample of objects.

\section{Conclusions}

For the first time we have built the overall nuclear SED of Centaurus
A, the only radio galaxy for which data provide a good coverage of the
nuclear emission from the radio to the $\gamma$--ray band. The SED
appears to be remarkably similar to that of blazars, as in a $\nu-\nu
L_\nu$ representation it is well represented by two broad peaks
located in the far-infrared band and at $\sim 0.1$ MeV.  Although we
cannot exclude that the emission observed at the various wavelengths
might be produced by different components/radiation processes, we have
shown that a simple one zone, homogeneous SSC model is adequate to
account for the overall spectral distribution. Note that in this
framework, the high IR polarization of the Centaurus A nucleus (Bailey
et al. 1986, Capetti et al. 2000) is readily explained as due 
to synchrotron emission.

In this SSC scenario we have found that the physical parameters 
(magnetic field, particles energy distribution etc.) are
in the range usually found for BL Lacs, with the noticeable exception
of a beaming factor lower than the values usually
derived from fitting the SED of BL Lacs ($\delta \sim 15-20$).  This is
qualitatively expected when a relativistic jet is seen at large angles
of sight.

However, while these results provide evidence that Centaurus A is
indeed a misoriented BL Lac and support the unification scheme for low
luminosity radio-loud AGNs, a more quantitative analysis shows that
the beaming factor derived for it, $\delta \sim 1$, is not simply
compatible with being originated in a misoriented fast ($\Gamma \sim
15-20$) blazar component.  In fact, for the viewing angle derived from
VLBI observations ($\sim 50^{\circ} - 80^{\circ}$), much smaller
values of $\delta$ would be expected. This suggests that the emission
from jets can be represented by a fast (highly relativistic) {\it
spine} and a slower {\it layer}, supporting the indications we have
found through the comparison of complete samples of BL Lacs and radio
galaxies (Chiaberge et al. 2000) and in agreement with other
independent observational evidence (e.g. Laing 1993, Laing et
al. 1998, Giovannini et al. 1999, Swain et al. 1998). The jet spine
dominates the emission in blazars and due to the large Lorentz
factors, its emission falls very rapidly as the angle of sight
increases.  Instead the slower layer would dominate in radio galaxies.
For a range of viewing angles (which depends on the details of the jet
velocity structure), the spine and layer emission can provide a
similar level of emission and thus be observed simultaneously.  If
these two components have significantly different spectral
properties (although this is not suggested by our analysis), the
resulting SED of these intermediate objects might be rather complex,
e.g. with multiple emission peaks.

Recent results suggest that the whole population of blazars follows
a sequence that links the overall spectral properties (in particular
the luminosity and location of the emission peaks) with the nuclear
radio luminosity, in turn related to the extended one.
Centaurus A nicely fits into this trend as by ``beaming'' its SED for
a typical BL Lac Doppler factor the frequency and the amplitude of the
peaks, as well as the (unbeamed) extended radio power, are typical of LBL.

Clearly the presence of significant emission from a layer has
important consequences also on the statistics of the sources, in
particular for the comparison of the counts and luminosity functions
of the beamed and parent populations. 
In these estimates, the role of the layer emission
can be identified with that played by some
'isotropic' emission, quantified by the relative fraction $f$ of
beamed and un--beamed luminosities (Urry \& Shafer 1984). The results
found for Centaurus A would correspond to $f\sim 1$. This is in
contrast with the much smaller values inferred by the modeling, as
required by both the relative small number of beamed sources (implying
high $\Gamma$) and the limits on the total luminosity range spanned by
the whole beamed and unbeamed populations.

The presence of a high energy peak in Centaurus A at $\sim 0.1$ MeV
raises the question of the contribution of radiogalaxies to the
gamma--ray background radiation as already suggested by Steinle et al.
(1998). However, if all radio-galaxies emit the same fraction of
bolometric luminosity as Centaurus A in this band, despite their much
larger number density, their contribution would be of only $\sim$ 10
per cent relative to that produced by blazars, because of the much
higher power radiated by beamed sources of similar extended radio
emission.

Simultaneous multiwavelength observations of a large sample of radio
galaxies, taking advantage of the future generation of $\gamma$--ray
instruments, will allow us both to confirm our present findings and to
establish if the trends observed in blazars occur also in their parent
population. This would provide strong support to the unification
models and important information on the physical structure of
relativistic jets.

\section*{Acknowledgments}

The authors acknowledge the Italian MURST for financial
support.

\bsp

\label{lastpage}


\begin{thebibliography}{}

\bibitem[00]{00} Barthel P. D. 1989, ApJ, 336, 606 

\bibitem[00]{00} Bailey J., Sparks W.\ B., Hough J.\ H.\ \& Axon
D.\ J.\ 1986, Nat, 322, 150

\bibitem[00]{00} Bloom S.\ D., Marscher A.\ P., 1993, in Friedlander M., 
Gehrels N., Macomb D.J., eds., Proc. CGRO AIP 280, New York, p. 578

\bibitem[00]{00} Burns J.\ O., Feigelson E.\ D.\ \& Schreier E.\
J.\ 1983, ApJ, 273, 128

\bibitem[00]{00} Capetti A., Trussoni E., Celotti A., Feretti L. \&
Chiaberge M. 2000, MNRAS, 318, 493

\bibitem[00]{00} Capetti A., et al. 2000, ApJ, 544, 269

\bibitem[00]{00} Cardelli J.\ A., Clayton G.\ C.\ \& Mathis, J.\ S.\
1989, ApJ, 345, 245

\bibitem[00]{00} Chiaberge M.\ \& Ghisellini G.\ 1999, MNRAS, 306,
551

\bibitem[00]{00} Chiaberge M., Capetti A., Celotti A., 1999, A\&A, 349, 77 

\bibitem[00]{00} Chiaberge M., Capetti A., Celotti A., Ghisellini G.,
2000, A\&A, 358, 104

\bibitem[00]{00} Dermer C.\ D.\ \& Schlickeiser R.\ 1993, ApJ, 416,
458

\bibitem[00]{00} Fossati G., Maraschi L., Celotti A., Comastri A.\
\& Ghisellini G.\ 1998, MNRAS, 299, 433

\bibitem[00]{00} Ghisellini G.\ \& Madau P.\ 1996, MNRAS, 280, 67

\bibitem[00]{00} Ghisellini G., Maraschi L.\ \& Dondi L.\ 1996,
A\&AS, 120, C503

\bibitem[00]{00} Ghisellini G., Celotti A., Fossati G., Maraschi L. \&
Comastri, A. 1998, MNRAS, 301, 451

\bibitem[00]{00} 
Giovannini G., Feretti L., Gregorini 
L.\ \& Parma P.\ 1988, A\&A, 199, 73 

\bibitem[00]{00} 
Giovannini G., 
Taylor G.\ B., Arbizzani E., Bondi M., Cotton W.\ D., Feretti L., 
Lara L.\ \& Venturi T.\ 1999, ApJ, 522, 101 

\bibitem[00]{00} Grandi P.\ et al.\ 2000, proceedings of 32nd COSPAR
Symposium (1998), Advances in Space Research, 25, 485

\bibitem[00]{00} Hawarden T.\ G., Sandell G., Matthews H.\ E.,
Friberg P., Watt G.\ D.\ \& Smith P.\ A.\ 1993, MNRAS, 260, 844

\bibitem[00]{00} Hui X., Ford H. C., Ciardullo R. \& Jacoby G.
H. 1993, ApJ, 414, 463

\bibitem[00]{00} Kellermann K.\ I., Zensus J.\ A.\ \& Cohen M.\ H.\
1997, ApJ, 475, L93

\bibitem[00]{00} Kinzer R.\ L.\ et al.\ 1995, ApJ, 449, 105 

\bibitem[00]{00} Laing R.A., 1993, In: Burgarella D., Livio M., O'Dea
C.P. (eds.) Space Telescope Sci. Inst. Symp. 6: Astrophysical
Jets. Cambridge University Press, Cambridge, p. 95

\bibitem[00]{00} Laing R.\ A., Parma P., de Ruiter H.\ R.\ \&
Fanti R.\ 1999, MNRAS, 306, 513

\bibitem[00]{00} Maraschi L., Ghisellini G.\ \& Celotti A.\ 1992,
ApJ, 397, L5

\bibitem[00]{00} Marconi A., Schreier E.\ J., Koekemoer A., Capetti
A., Axon D., Macchetto D.\ \& Caon N.\ 2000, ApJ, 528, 276

\bibitem[00]{00} Marconi A., Capetti. A, Axon D.J., Koekemoer A.,
Macchetto F.D., Schreier E.J., 2001, ApJ, in press, astro-ph/0011059

\bibitem[00]{00} Mirabel I.\ F.\ et al.\ 1999, A\&A, 341, 667

\bibitem[00]{00} Padovani P.\ \& Giommi P.\ 1995, ApJ, 444, 567

\bibitem[00]{00} Rothschild R.\ E.\ et al.\ 1999, ApJ, 510, 651

\bibitem[00]{00} Sikora M., Begelman M.\ C.\ \& Rees M.\ J.\ 1994,
ApJ, 421, 153

\bibitem[00]{00} Steinle H. et al. 1998, A\&A, 330, 97

\bibitem[00]{00}
Swain M.\ R., 
Bridle A.\ H.\ \& Baum S.\ A.\ 1998, ApJ, 507, L29 

\bibitem[00]{00} Tingay S.\ J.\ et al.\ 1998, AJ, 115, 960

\bibitem[00]{00} Urry C.M., Padovani P., 1995, PASP, 107, 803

\bibitem[00]{00} Urry C.\ M.\ \& Shafer R.\ A.\ 1984, ApJ, 280, 569


\end{thebibliography}
\end{document}